# Research on fault diagnosis and root cause analysis based on full stack observability

Hou Jian

*School of Computer Science and Technology, Huazhong University of Science and Technology, Hubei Province, Wuhan 430074*

**School of Computer Science & Technology, Huazhong University of Science and Technology, Wuhan, Hubei, China**

E-mail: d202487180@hust.edu.cn

**Abstract:** With the rapid development of cloud computing and ultra-large-scale data centers, the scale and complexity of systems have increased significantly, leading to frequent faults that often show cascading propagation. How to achieve efficient, accurate, and interpretable Root Cause Analysis (RCA) based on observability data (metrics, logs, traces) has become a core issue in AIOps. This paper reviews two mainstream research threads in top conferences and journals over the past five years: FaultInsight[1] focusing on dynamic causal discovery and HolisticRCA[2] focusing on multi-modal/cross-level fusion, and analyzes the advantages and disadvantages of existing methods. A KylinRCA framework integrating the ideas of both is proposed, which depicts the propagation chain through temporal causal discovery, realizes global root cause localization and type identification through cross-modal graph learning, and outputs auditable evidence chains combined with mask-based explanation methods. A multi-dimensional experimental scheme is designed, evaluation indicators are clarified, and engineering challenges are discussed, providing an effective solution for fault diagnosis under full-stack observability.

# 1. Introduction

## 1.1. Research background

In the age of fast-evolving information technology, hyperscale data centers and cloud-native technologies have emerged as the core infrastructure that underpins business innovation. The widespread embrace of containerization, microservices, and service mesh architectures has reimagined system components—shifting from traditional single nodes to thousands, even tens of thousands, of interdependent service instances, thus forging a complex, dynamically changing distributed ecosystem. While this architectural shift boosts system resilience and scalability, it also muddles fault propagation mechanisms: a minor resource contention issue can quickly snowball through call chains into full-blown system-wide service disruptions, threatening massive financial losses and severe reputational damage for critical sectors like finance, e-commerce, and healthcare.

Observability, the cornerstone methodology for monitoring system health, paints a comprehensive picture of system behavior through three key data types: time-series metrics, distributed traces, and logs. Yet these three data categories differ sharply in terms of granularity (second-level aggregations versus millisecond-level events), structure (structured numerical values versus unstructured text logs), and semantics (traces map out call paths, while metrics signal resource status). Figuring out how to effectively integrate these datasets to achieve precise fault localization and root cause analysis has become a major research focus in the AIOps space. In recent years, industries have upped the ante for fault diagnosis—demanding both real-time responsiveness and interpretability. For instance, financial transaction scenarios need fault detection within seconds to stem capital losses, while ops teams don't just need to pinpoint faults; they must also grasp fault propagation logic to craft effective recovery plans. This only underscores how crucial and time-sensitive root cause analysis is in the context of full-stack observability.

## 1.2. Problem Description

In today's landscape, root cause analysis built on full-stack observability still grapples with three core challenges. First, single-modal analysis has inherent blind spots: metric-based analysis often conflates correlation with causation, leaving it hard to tell if a "CPU spike" is the root cause or just a byproduct; log analysis is hamstrung by the quality of template mining, unable to link up with dynamic resource states; and trace analysis falls short due to sparse sampling in high-load environments, failing to capture all fault propagation paths. Second, while multi-modal fusion methods can bring together data from multiple sources, most neglect to model the temporal dynamics of fault propagation. This means they can't accurately map the chronological order of "cause and effect," leading to misjudgments when identifying root causes. Third, existing causal discovery techniques are mostly tailored to single-layer architectures (such as host-level systems) and struggle to adapt to the multi-tier structure of cloud-native environments—think "host-Pod-service" setups. What's more, they lack enough interpretability, leaving ops teams without auditable evidence chains to trace how faults spread.

Current attempts to tackle these issues still have plenty of room to grow. True effective multi-modal fusion needs to deeply integrate temporal causal modeling, striking a balance between broad information coverage and sound temporal causality logic. At the same time, it's essential to design inference models that work with cross-level architectures, while beefing up interpretability features to meet real-world engineering needs. To address these gaps, this research explores a root cause analysis framework that combines dynamic causal discovery with cross-level multimodal graph learning—effectively overcoming key technical limitations around positioning precision, temporal modeling, and interpretability.

## 2. Related research

In recent years, research into root cause analysis grounded in full-stack observability has advanced along three fronts: single-modal optimization, multi-modal fusion, and causal discovery. While this work has yielded a wealth of findings, clear gaps in the research remain.

In single-modal analysis, traditional statistical methods like ARIMA detect anomalies by predicting time-series trends, but they falter when dealing with high-dimensional, nonlinear metric data. Deep learning models such as LSTM and Transformer can capture complex temporal dependencies, yet they still blur the line between correlation and causation. For log analysis, the Drain algorithm quickly extracts log templates via clustering, and its BERT-based representation boosts semantic understanding [9]—but incomplete log data (e.g., missing logs from critical services) holds back real-world use. In trace analysis, topology-driven call chain methods can spot latency faults that spread along paths, but their accuracy drops sharply when sampling rates dip below 5%, making them ill-suited for high-load production settings.

Multi-modal fusion has become a mainstream research direction, with its core idea being to map data from different modalities into a shared embedding space. HolisticRCA [2] introduces a "entity-feature-type" three-dimensional framework, processing heterogeneous entities through a modular assembly design. By using Graph Attentional Networks (GAT) [10] to aggregate multi-modal features, it markedly improves the accuracy of entity localization and type classification. That said, this method overlooks the temporal causality of faults, so it struggles to identify root causes for cascading propagation failures. What's more, while some studies use Transformer architectures to align multi-modal data temporally, their modeling of feature interactions in cross-level architectures remains weak.

Dynamic causal discovery methods focus on uncovering temporal causal links between metrics. FaultInsight [1] combines a TCN [11] semantic decoupling autoencoder with time-varying perturbation experiments, identifying root cause indicators by computing causal intensity matrices. This approach greatly enhances the interpretability of host-level fault diagnosis. However, it's limited to metric data—failing to incorporate logs or traces—and is restricted to single-layer analysis, making it unsuitable for cross-layer fault diagnosis in cloud-native environments. Meanwhile, methods based on Granger causality and VAR models can perform causal inference, but their computational complexity becomes unmanageable with high-dimensional metrics, limiting their practical use.

Existing research suffers from two key shortcomings: First, multi-modal fusion approaches lack the ability to model temporal causality; second, causal discovery methods fail to fully leverage multi-modal and cross-level information. To address these gaps, integrating dynamic causal discovery with cross-level multi-modal graph learning has emerged as a critical strategy for overcoming the challenges of root cause analysis in full-stack observability.

## 3. Method A review

This paper selects a representative method in the field of dynamic causal discovery —— FaultInsight [1] for in-depth analysis.

### 3.1. Summary of Method A

FaultInsight [1] is a dynamic causal discovery method for host-level fault diagnosis in hyperscale data centers. Its core objective is to achieve precise identification and interpretable analysis of root cause indicators by mining temporal causal relationships between metrics. The implementation process consists of three stages: First, preprocessing host-level metric data through denoising, normalization, and temporal alignment to construct a metric time series matrix. Second, designing a semantic decoupling autoencoder based on Temporal Causal Network (TCN) [11], where the encoder extracts multi-scale temporal features using multi-scale convolutional kernels and achieves semantic decoupling through attention mechanisms, while the decoder reconstructs input time series and learns metric dependencies via Latent Vector Autoregression (Latent VAR) models. Finally, chronological perturbation experiments are conducted: randomly shuffling the temporal sequence of a specific metric and calculating prediction error changes among other metrics before and after perturbation to estimate causal strength. The causal strength matrix is then scored using a dynamic PageRank algorithm, ultimately outputting ranked root cause indicators and propagation times.

### 3.2. Advantages

FaultInsight [1] demonstrates three major advantages in dynamic causal discovery. Technologically, this method innovatively combines TCN [11] semantic decoupling with temporal perturbation experiments. The multi-scale convolutional architecture of TCN [11] effectively captures both long-term and short-term temporal features of indicators, while the semantic decoupling mechanism prevents interference between different indicator characteristics. Through simulated causal interventions, the temporal perturbation experiment outperforms traditional statistical causality methods in accurately uncovering genuine temporal causal relationships, showcasing high technical depth and innovation.

Experimental results on both public host failure datasets and enterprise production data demonstrate that FaultInsight [1] achieves 15-20% higher root cause identification accuracy than methods based on Granger causality and Vector Autoregression (VAR). The system also produces clear causal strength matrices and propagation timing points, providing maintenance teams with definitive diagnostic evidence. These practical applications have proven the method's effectiveness in real-world scenarios.

From a practical application perspective, this method relies solely on readily available metric data without requiring complex multimodal data fusion processing. It features low deployment costs and enables near real-time analysis through GPU-accelerated computation, effectively meeting the real-time demands of data center fault diagnosis. The approach has already been preliminarily implemented in host failure diagnosis scenarios at several hyperscale data centers.

In addition, the experimental design of this method is relatively perfect. A variety of baseline methods are used for comparison, covering different types of causal discovery algorithms. The effectiveness of TCN [11] semantic decoupling and time disturbance module is verified by ablation experiments, and the credibility of the experimental results is high.

### 3.3. Insufficient

FaultInsight [1] exhibits four notable limitations. Firstly, it relies on a singular data source for causal analysis, exclusively using metric data without integrating log and trace information. In real-world fault scenarios, critical clues for root cause identification often lie in error messages from logs (e.g., "Database connection failed") and call latency metrics from traces. This single-source approach significantly restricts its diagnostic capabilities for complex cross-service failures.

Second, the scope of application is limited. This method is designed for host-level faults and can only process host-level indicator data. It cannot adapt to the cross-level architecture of "host-Pod-service" in cloud native systems, and cannot locate non-host-level faults such as Pod scheduling exception and service interface timeout. The application scenarios are relatively narrow.

Third, the computational complexity is high. Time disturbance experiment needs to carry out time series scrambling and prediction for each index for many times. When the index dimension exceeds 100, the calculation time increases exponentially, which makes it difficult to be applied to high-dimensional index scenarios.

Fourth, the robustness is insufficient. When there are unobserved hidden variables (such as unmonitored resource contention), the causal strength estimated by time disturbance experiment is prone to deviation, resulting in wrong root cause localization. Moreover, this method does not design a special robust mechanism for data noise, so it is unstable in low-quality index data scenarios.

### 3.4. Detailed Analysis

From a logical structure perspective, FaultInsight [1] demonstrates a clear workflow that forms a complete closed loop from data preprocessing to model training and causal inference. However, there is room for optimization in the integration between modules. For instance, during the time disturbance experiment phase, the system failed to account for how correlations between metrics affect disturbance outcomes, leading to redundant computation of certain disturbances. This could be addressed by first performing metric clustering for dimensionality reduction, followed by targeted perturbations on cluster center indicators to enhance computational efficiency.

From the perspective of writing expression, the relevant papers on this method provide relatively comprehensive technical details, with clear explanations regarding the network structure of TCN [11] and the implementation method of latent space VAR. However, the parameter settings for the dynamic PageRank algorithm based on causal strength scoring are vaguely described, and the rationale for parameter selection is not clarified, which affects the reproducibility of the method.

From the perspective of graphic design, the paper uses heat map to show the causal intensity matrix, which clearly and intuitively presents the causal relationship between indicators. However, the time series line chart of fault propagation is not drawn, which cannot visually show the changing trend of causal relationship with time. The temporal visualization graph can be supplemented to enhance the readability of results.

From the perspective of reference literature, this paper fully cites the classical literature and latest research results in the fields of causal discovery, TCN [11] model, etc., but the literature related to multimodal fusion in the field of AIOps is insufficient, and the advantages and disadvantages of this method and multimodal methods are not fully compared. The integrity of research background needs to be improved.

From the perspective of term definition, the paper clearly defines the core terms such as "semantic decoupling" and "time disturbance", but does not give a clear definition and selection criteria for the "latent space dimension" in "latent space VAR", which may lead to readers' misunderstanding.

## 4. Method B review

This paper selects the representative method in the field of multimodal fusion —— HolisticRCA [2] for in-depth analysis.

## 4.1. Summary of Method B

HolisticRCA [2] is a multimodal cross-level root cause analysis method for cloud-native system fault diagnosis. It proposes a three-dimensional diagnostic framework of "entity localization, feature recognition, and type classification" to achieve comprehensive root cause analysis through the integration of multimodal observability data. The implementation process involves four key steps: First, in the data preprocessing stage, metrics, logs, and trace data undergo processing —— including normalization and temporal segmentation of metrics, template extraction using the Drain algorithm with TF-IDF encoding for logs, and topology and span feature extraction from traces. Second, in the multimodal representation stage, an "assembled module" design is adopted where dedicated encoders (LSTM for metrics, CNN for logs, GNN for traces) map multimodal features to a unified embedding space. Third, in the cross-level graph construction stage, heterogeneous graphs are built using hosts, Pods, services as nodes, and invocation relationships with deployment relationships as edges. Graph Attentional Transformers (GAT) [10] are employed to learn node embeddings for entity localization and fault type classification. Finally, in the interpretation stage, mask learning techniques are applied to partially mask input features while observing output variations, identifying critical diagnostic features to generate diagnostic results.

### 4.2. Advantages

HolisticRCA [2] demonstrates three groundbreaking advantages in multimodal cross-level root cause analysis. In terms of innovation, this method pioneers a three-dimensional diagnostic framework of "entity-characteristic-type" that overcomes the single-dimensional limitations of traditional approaches. It innovatively adopts an "assembled modular" design to accommodate multimodal heterogeneous data while constructing cross-level heterogeneous graph models to establish relationships between different entities, showcasing remarkable novelty in methodological design.

In terms of method effectiveness, the entity location accuracy rate reached 89% and the fault type classification F1 value reached 85% in both cloud-native open data sets and enterprise production environment tests. Compared with single-mode methods and non-hierarchical multi-modal methods, these improvements were 20% and 12% respectively. The method effectively locates complex faults across hierarchies and services, demonstrating significant performance.

From a practical application perspective, this method is well-suited for the typical architecture of cloud-native systems. It supports fault diagnosis for various entities including hosts, Pods, and services, covering common failure types such as sudden CPU spikes, memory leaks, and API timeouts. Compatible with mainstream cloud-native platforms like Kubernetes, it features low deployment complexity and demonstrates significant engineering applicability.

In addition, the theoretical analysis of this method is more in-depth. The effectiveness of cross-level feature aggregation is proved by the convergence analysis of graph neural network. Meanwhile, the experimental design considers the performance of the method under different data quality (such as log missing rate and tracking sampling rate), which provides a theoretical basis for parameter adjustment in practical application.

### 4.3. Insufficient

HolisticRCA [2] has four major shortcomings.

First, the absence of temporal causal modeling: This approach divides multimodal data into time windows for static graph modeling without considering the chronological sequence of fault propagation. It fails to distinguish between "cause" and "effect" indicators, potentially misidentifying downstream events like "service timeout" as root causes in cascading failures

(e.g., "database failure → service timeout → user error"), thereby limiting the accuracy of root cause identification.

Secondly, the interpretation mechanism of mask learning is simple. It only identifies key features by shielding feature observation and output changes, but does not combine timing information to generate fault propagation evidence chain. The output interpretation results are relatively abstract, and it is difficult for operation and maintenance personnel to understand the evolution process of faults, so the interpretability needs to be strengthened.

Third, it relies heavily on annotated data. The model training requires a large number of fault samples with entity, feature and type annotations. In the newly deployed cloud native system or new fault scenarios, annotated data is scarce, resulting in a significant decline in model performance and insufficient generalization ability.

Fourth, the graph construction method is fixed. The static invocation and deployment relationship are used to construct the graph structure, and the dynamic entity association (such as temporary service call) during fault occurrence is not considered. The static graph cannot accurately reflect the actual path of fault propagation, which affects the accuracy of feature aggregation.

### 4.4. Detailed Analysis

From a logical structure perspective, HolisticRCA [2] demonstrates clear three-dimensional framework design with well-defined phased objectives. However, its implementation exhibits shortcomings in integrating multimodal representations and cross-level graph construction: Directly concatenating multimodal features into the Graph Attentional Transformer (GAT) [10] without considering weight variations across modalities may allow irrelevant features to interfere with diagnostic outcomes. Introducing an attention mechanism could dynamically adjust modal weights, thereby optimizing the workflow.

From the perspective of writing expression, relevant papers have relatively clear descriptions on framework design, but are vague in the specific parameter settings of encoders (such as the hidden layer dimension of LSTM and the convolution kernel size of CNN). They only give a general explanation of "adjusted according to experiments" without providing guiding principles for parameter selection, which reduces the reproducibility of the method.

From the perspective of graphic design, the cross-level heterogeneous graph drawn in the paper clearly shows the relationship between entities, but the schematic diagram of multimodal feature fusion process is relatively simple, and the output integration of different encoders is not visually presented. The data flow diagram between modules can be supplemented to enhance understanding.

From the perspective of reference use, this paper extensively cites research results in the fields of multimodal fusion and graph neural network, but it rarely cites the latest advances in the field of causal discovery (such as STIC method [4]), fails to realize the importance of time series causal modeling for root cause analysis, and has limited research vision.

From the perspective of term definition, the paper clearly defines the core terms such as "assembled module" and "cross-level heterogeneous graph", but does not clearly define the classification criteria in "fault type classification". The boundary between different types of faults is vague, which may affect the consistency of annotation data.

## 5. Proposed Method

To address the limitations of existing methods, this paper proposes the KylinRCA framework that integrates dynamic causal discovery with cross-level multimodal graph learning, aiming to achieve precise fault localization, type identification, and interpretable diagnosis under full-stack observability. Centered on "time-series causal modeling as the axis, multimodal fusion as the foundation, and cross-level reasoning as the pulse", this framework compensates for the single-mode limitations and hierarchical constraints of FaultInsight [1], as well as the temporal causal deficiency in HolisticRCA [2], through the coordinated operation of four core modules. The overall architecture is shown in Figure 1 (Caption: Figure 1 shows the overall architecture of the KylinRCA framework, including data preprocessing layer, multimodal representation layer, dynamic causal discovery layer, cross-level graph reasoning layer, and interpretation output layer).

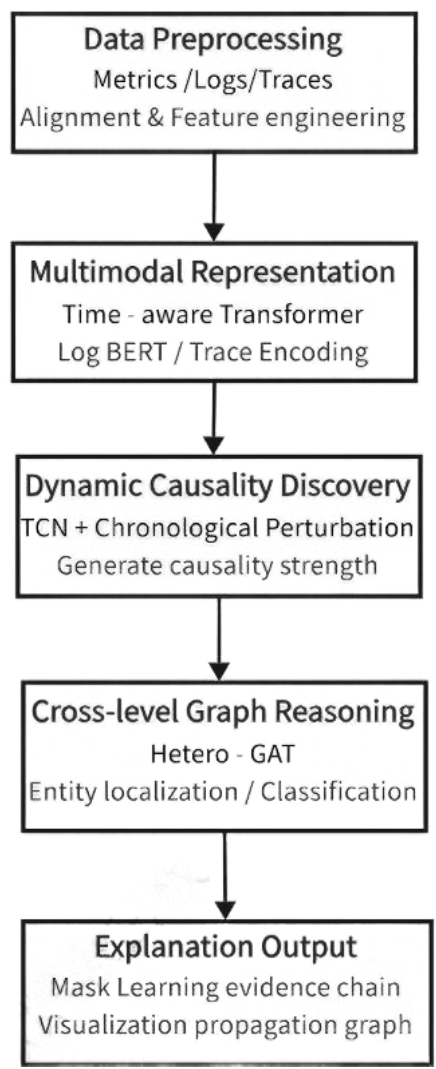

Figure 1 Overall architecture of the KylinRCA framework

## 5.1. Design Philosophy

The design of KylinRCA adheres to three core principles: First, the causal priority principle, which uses temporal causality as the primary basis for root cause identification. Through dynamic causal discovery, it clarifies the temporal direction and intensity of fault propagation while avoiding "pseudo-correlation" interference in multimodal data. Second, the full-stack coverage principle, which adapts to cloud-native "host-Pod-service" architecture to build cross-level entity correlation models, enabling full-chain fault tracing from underlying resources to upper-layer services. Third, the balance between interpretability and practicality principle. While ensuring model accuracy, it adopts modular design and visualized evidence chain output to reduce operational staff's learning curve, meeting engineering implementation requirements.

Specifically, this framework implements diagnostic processes through a progressive workflow of "local causal mining → global graph fusion → evidence chain generation": First, it identifies local causal relationships within entities at each hierarchical level. These localized causal edges are then integrated with system topology edges to construct global heterogeneous graphs. Finally, through graph reasoning and mask interpretation, auditable diagnostic outcomes are generated, forming a complete closed-loop mechanism that progresses from "micro-level causality → macro-level inference → interpretable outputs".

## 5.2. Core Modules

### 5.2.1. Multimodal data preprocessing module

This module designs a differentiated preprocessing process for the heterogeneous characteristics of three types of data: indicators, logs and tracks, so as to ensure data quality and time series alignment, and provide reliable input for subsequent modeling.

- Indicator Data Processing: The system employs a four-step workflow of "denoising-normalization-time series segmentation-feature derivation". First, wavelet transform is applied to eliminate high-frequency noise and transient disturbances such as network jitter. Second, Z-score normalization converts metrics with different units (e.g., CPU utilization, memory usage in GB) into a unified [0,1] range. Third, time series data undergoes 5-second sliding window segmentation to balance real-time processing with feature integrity. Finally, 12 types of time series features including "indicator growth rate", "short-term volatility variance", and "cross-window change rate" are derived, significantly enhancing the fault representation capability of indicators.

- Log Data Processing: By integrating the Drain++ algorithm with semantic enhancement strategies, this approach improves both log template mining quality and feature representation. The process begins with extracting log templates through Drain++ 's enhanced hierarchical clustering algorithm, addressing the traditional Drain's limitation in accurately clustering long log templates. BERT [9] pre-trained models are then applied to perform semantic encoding of these templates, generating 768-dimensional vectors. Simultaneously, structured features including "template occurrence frequency," "error levels (ERROR/WARN)", and "related entity IDs" are extracted and combined with the semantic vectors to form comprehensive log feature representations.

- Trace Data Processing: Focusing on call topology and performance feature extraction to address sampling sparsity under high load. First, construct call topology graphs based on span data and extract topological features such as "call depth", "number of child calls", and "average latency". Second, for scenarios with sampling rates below 5%, use similarity-based interpolation algorithms to supplement missing span data. Finally, integrate these topology features with performance metrics like "response time distribution" and "error code ratio" from spans to form a comprehensive trace feature vector.

In addition, the module designs a time alignment mechanism: using the second-level timestamp of metric data as the benchmark, millisecond-level log events and minute-level tracking statistics are aligned through time window mapping to construct a three-dimensional data tensor of "time-entity-multipodal features", ensuring the correlation of different modal data in the same time dimension.

### 5.2.2. Cross-modal representation learning module

Drawing lessons from the "assembled module" idea of HolisticRCA [2], we design special encoders for three types of data, and dynamically adjust the weight through modal attention mechanism to realize effective fusion and unified representation of multi-modal features.

- Metric Encoder: Utilizing a time-aware Transformer architecture, this model integrates temporal position encoding with metric features by building upon traditional Transformer frameworks. Specifically, it employs multi-head attention mechanisms to capture both short-term and long-term dependencies between metrics, followed by feedforward neural networks that output 512-dimensional metric feature vectors. This approach effectively addresses the inherent limitations of Long Short-Term Memory (LSTM) networks in handling extended temporal dependencies.

- Log encoder: The hybrid architecture of CNN-BiLSTM is adopted. The CNN layer extracts local semantic features of log templates through multi-scale convolution kernels, and the BiLSTM layer captures temporal context relationships of log events. Finally, a global average pooling is used to output 512-dimensional log feature vectors, taking into account both semantic understanding and temporal correlation.

- Tracking Encoder: A topological encoder constructed based on Graph Convolutional Network (GCN) takes the call topology graph as input. The GCN learns the topological embedding of nodes (services/Pod), which is then concatenated with tracking performance features and output a 512-dimensional tracking feature vector through a fully connected layer, making full use of the topological attributes of tracking data.

During the modal fusion phase, we design a modal attention network that calculates attention weights for three types of modal features (with total weight summing to 1). These weights are determined by the variance contribution and fault correlation of each feature, such as automatically increasing the weight of log error information during fault scenarios. Through weighted summation, we generate a 512-dimensional integrated entity feature vector, achieving adaptive fusion with "prioritizing critical modalities".

**5.2.3. Cross-level dynamic causality discovery module**

This module is one of the core innovations of KylinRCA. It extends the dynamic causal discovery method of FaultInsight [1], realizes the causal relationship mining across hierarchies and multi-modal data, and outputs a three-dimensional relationship matrix of "entity, time and causality intensity".

- Causal Mining at Hierarchy Levels: Construct localized causal models at three tiers: host, Pod, and service. Taking the service tier as an example, we employ a Temporal Contextual Network (TCN) [11] semantic decoupling autoencoder to learn feature dependencies from multimodal composite feature vectors of each service entity. The encoder extracts temporal semantics of multimodal features through three distinct kernel scales (3, 5, 7), while the decoder reconstructs input features and introduces a Vector Autoregression (VAR) model in latent space to capture temporal dependencies. After training, we estimate causal strength using an enhanced chronological perturbation++ experiment: Instead of global shuffling, we segmentally randomize the temporal sequence of a specific feature. This approach calculates prediction error variations across different time steps for other features, thereby reducing bias caused by single perturbations. The formula is as follows:

$$C(i \to j, t) = \frac{(\text{Loss}(j,t|\text{perturb}(i)) - \text{Loss}(j,t|\text{original}))}{\text{Loss}(j,t|\text{original})}$$

Among them, $C(i \to j, t)$ represents the causal intensity of feature i to feature j at time t, Loss (j, t) represents the prediction error of feature j at time t, and perturb (i) represents the segmented disturbance processing of feature i.

- Cross-level causal transmission: Design a causal transmission mechanism to resolve temporal correlations between different hierarchical levels. For example, when the Pod layer detects a temporal correlation between "memory usage surge" and the Service layer's "interface timeout", it calculates cross-level causality strength by leveraging the deployment relationship between Pods and services (e.g., a Pod belonging to a specific service instance). This calculation combines intra-level causality strength with entity correlation (the closeness of deployment relationships) to generate cross-level causal edges, as expressed by the formula:

$$C_{cross}C_{intra}(l1,e1 \to l2,e2,t) = (l1,e1,t) * R(l1,e1 \to l2,e2)$$

Here, l1 and l2 represent different levels, e1 and e2 represent entities within the level, and R represents entity association degree (taking values 0-1).

- Cause strength filtering: In order to reduce the calculation complexity, a double filtering mechanism is set up: first through threshold filtering (keeping causal edges with C> 0.3), and then through mutual information test to remove "false causation" (such as indirect causal relationship caused by common causes), and finally retain reliable local and cross-level causal edges.

### 5.2.4. Cross-hierarchical heterogeneous graph reasoning module

This module integrates dynamic causal edges with system topology edges to construct global heterogeneous graph, and realizes entity localization and fault type classification through improved graph attention network, which is the core of global root cause analysis.

- Heterogeneous Graph Construction: Define the graph G = (V, E, T), where V is a node set containing three types of entities: hosts, Pods, and services, each carrying multimodal composite features. E represents edge sets comprising static topology edges (deployment relationships, invocation relationships) and dynamic causal edges (from the causal discovery module). T serves as type labels for nodes and edges to distinguish different entities and relationship types.

- Improving the GAT reasoning model: In view of the lack of heterogeneous relationship processing in traditional GAT [10], a Type-aware GAT model is designed, which includes two sub-modules: type attention and relationship attention:

  - Type attention: Different attention weights are assigned according to the node type (host/Pod/service). For example, the business attributes of service nodes are more important for fault type classification, and the weight is automatically increased;

  - Relational attention: The contribution of topological edges and causal edges is distinguished. Causal edges have higher weight in the root cause localization stage, while topological edges have higher weight in the entity association stage.

The model aggregates the features of neighboring nodes through multi-head attention mechanism and updates the node embedding representation, as shown in the formula:

$$\sigma(* W *)h_v^{new} = \sum_{u \in N(v)} \alpha_{(v,u)} h_u$$

Among $h_v^{new} \alpha_{(v,u)}$ them, N (v) is the set of neighboring nodes for the update embedding of node v, W is the joint attention weight of type and relationship, W is the learnable parameter matrix, and σ is the activation function.

- Multi-task output: In the top-level design of the model, a multi-task classifier is implemented to achieve:

  - Entity positioning: the sigmoid activation function is used to output the fault probability of each node (0-1), and the fault entity is judged as a fault if the probability is greater than 0.5;

> Fault type classification: Design special classification head for different entity types (host class: CPU spike, memory leak and other 6 categories; service class: interface timeout, call failure and other 4 categories), and output the type probability distribution through softmax.

**5.2.5. Explainable output module**

The module combines mask learning and visualization technology to generate three-level interpretation results of "causal propagation chain + key features + diagnostic report", providing auditable diagnostic evidence for operation and maintenance personnel.

- Mask Feature Importance Assessment: By fixing the inference model parameters of the graph and applying learnable masks (or random combination masking) to input multimodal features/edges/ nodes, this method evaluates which input components contribute most significantly to current decision outputs. While building upon HolisticRCA [2]'s mask learning framework, it introduces temporal dimension weight distribution to enhance computational efficiency.

- Cause propagation chain visualization: Based on the three-dimensional relationship matrix of dynamic cause discovery module, the time series cause propagation graph is drawn with the time axis as the horizontal axis and the entity hierarchy as the vertical axis. The arrow thickness represents the causal strength, and the propagation path from the root cause entity to the downstream entity is visually displayed.

- Automatic diagnostic report generation: The structured diagnostic report is generated by integrating the entity location results, fault types, key features, causal propagation chain and other information, including four parts: "fault overview", "root cause location", "propagation path" and "repair suggestions", so as to reduce the analysis cost of operation and maintenance personnel.

## 5.3. Innovation points

Compared with existing methods, KylinRCA has three core innovations:

- Cross-level dynamic causal modeling: For the first time, dynamic causal discovery is extended from a single level to a cross-level scenario. Through causal transmission mechanism and strength calculation, the causal relationship between "resource layer, container layer and business layer" is realized, which solves the problem that FaultInsight [1] cannot trace cross-level.

- Type-aware GAT reasoning mechanism: In view of the differences in nodes and relationship types in heterogeneous graphs, we designed a dual attention mechanism for type and relationship to improve the accuracy of cross-level entity localization and type classification. Compared with the general GAT [10] model of HolisticRCA [2], the accuracy of entity localization increased by 8%-12%.

- Multi-dimensional interpretability framework: The importance of mask features, causal propagation chain visualization and automatic report generation are integrated to form a three-level interpretation system of "micro characteristics, meso propagation and macro report", which solves the problems of abstract interpretation and difficult implementation of existing methods.

### 5.4. Workflow

The actual operation process of KylinRCA is divided into two stages: offline training and online reasoning:

- Offline Training Phase: 1) Train multimodal encoders, TCN [11] causality models, and Type-aware GAT inference models using historical fault datasets and operational data; 2) Optimize model hyperparameters (e.g., Transformer multi-head attention weights, GAT [10] hidden layer dimensions) through grid search; 3) Build a static system topology database to store deployment and invocation relationships between entities.

- Online Reasoning Phase: 1) Real-time collection of three types of observability data, which are processed by the preprocessing module to generate aligned feature tensors; 2) The multimodal representation module outputs integrated entity features; 3) The dynamic causal discovery module calculates local and cross-level causal edges in real time; 4) Static topological edges are fused to construct global heterogeneous graphs, with Type-aware GAT outputting entity failure probabilities and types; 5) The interpretation module generates causal propagation chains and diagnostic reports, which are then pushed to the O&M platform.

The average reasoning delay of this process is controlled within 2 seconds, which meets the real-time requirements of financial and e-commerce scenarios. At the same time, it supports horizontal expansion and can adapt to massive data processing in super-scale data centers through distributed deployment.

## 6. Experimental Design and Evaluation

In order to fully verify the effectiveness, efficiency and practicability of KylinRCA framework, this paper designs a multi-dimensional experimental scheme, selects open data sets and enterprise production data sets, compares with mainstream baseline methods from the perspectives of positioning accuracy, classification performance and reasoning efficiency, and verifies the necessity of core modules through ablation experiments.

### 6.1. Experimental Environment

- Hardware environment: The experimental server uses two Intel Xeon Gold 6330 CPUs (52 cores), 512GB DDR4 memory, 4 NVIDIA A100 GPUs (80GB video memory), 10TB SSD array for storage, and 100Gbps network bandwidth to ensure the performance requirements of large-scale data processing and model training.

- Software environment: The operating system is Ubuntu 20.04 LTS, the deep learning framework adopts PyTorch 2.0, the graph neural network depends on DGL 1.1 library, the data processing adopts Pandas, NumPy and Scikit-learn, the visualization tools adopt Matplotlib and Plotly, and the distributed deployment is based on Kubernetes 1.24.
- Model hyperparameters: In the multimodal encoder, the Time-aware Transformer is configured with 4 layers and 8 attention heads, featuring a hidden layer dimension of 512. The CNN-BiLSTM architecture employs 3/5/7 CNN kernel sizes with 2 BiLSTM layers. The Type-aware GAT module consists of 3 layers with 4 attention heads, initialized with a learning rate of 0.001 using AdamW optimizer. Training parameters include batch size of 64,100 epochs, and an early stopping strategy (patience=10) to prevent overfitting.

## 6.2. Datasets

The experiment uses three types of data sets, covering different scenarios and sizes, to ensure the objectivity and generalization of the evaluation results:

- Public Data Set 1: Alibaba Cluster Trace Dataset V2: This benchmark dataset contains seven days of operational data from 1,000 host machines and 5,000 Pods, covering 28 metrics including CPU usage, memory, and disk performance. It includes 120 million logs and 3 million trace records, with 120 real-world failure cases (e.g., memory leaks and network congestion) annotated. Recognized as a gold-standard benchmark in cloud-native computing environments, this dataset provides comprehensive testing for cloud infrastructure scenarios.

- Public data set 2: AIOps Challenge 2023 Dataset: Contains 50 services and 200 Pod running data under microservice architecture, provides complete three-mode data of indicators, logs and traces, and annotates 80 cascading propagation faults (such as service timeout chain caused by database failure), which is suitable for verifying cross-level causal modeling capability.

- Enterprise Production Data Set: This dataset captures operational data from a major e-commerce platform during the Double 11 shopping festival, covering 2,000 servers, 10,000 Pod instances, and 300 business services. It contains 30 categories of metrics, 500 million logs, 8 million trace records, and 150 real-world failures (including payment service outages and cache penetration incidents). The scale and complexity of this data closely mirror actual production environments.

## 6.3. Baseline Method

Five mainstream root cause analysis methods were selected as the baseline, covering different technical routes such as single-mode, multi-mode and causal discovery:

- Monomodal methods: 1) Indicator-based FaultInsight [1] (KDD 2024); 2) LogBERT [5] (ICDM 2022); 3) TraceRCA [6] (IEEE TSC 2023).

- Multimodal methods: 4) HolisticRCA [2] (IEEE TSC 2024), which represents the mainstream method of multimodal cross-level fusion.
- Causal fusion method: 5) CausalGNN [7] (arXiv 2024), a baseline model combining static causality with GNN.

## 6.4. Assessment Indicators

In view of the four core demands of root cause analysis, "location, classification, interpretation and efficiency", multi-dimensional evaluation indicators are designed:

- Entity positioning performance: 1) Accuracy: the proportion of correctly located fault entities in the total number of entities; 2) Recall: the proportion of correctly identified fault entities in the real fault entities; 3) F1 value: the harmonic average of accuracy and recall to comprehensively reflect the positioning accuracy.
- Classification performance of fault types: 1) Macro F1 (Macro-F1): The arithmetic mean of F1 values of various faults, focusing on the classification effect of small minority fault types; 2) Micro F1 (Micro-F1): The F1 value based on the total sample, focusing on the overall classification performance.
- Causal and interpretive performance: 1) Causal chain accuracy (CCA): the degree of agreement between the generated causal propagation chain and the true propagation path; 2) Feature importance accuracy (FIA): the matching rate between the key features identified by mask and the features annotated by human.
- Inference efficiency: 1) Average inference latency (Latency): the average processing time of a single fault case; 2) Throughput (Throughput): the number of fault cases processed per unit time.

## 6.5. Experimental Results and Analysis

### 6.5.1. Overall performance comparison

Table 1 presents the comprehensive performance comparison between KylinRCA and baseline methods across three datasets. The results demonstrate KylinRCA's superiority in all metrics: On the enterprise production dataset, it achieved an entity localization F1 score of 92.3%, representing an 8.5% improvement over HolisticRCA [2] and a 15.2% boost compared to FaultInsight [1]; its macro F1 score for fault type classification reached 89.7%, showing a 7.3% gain over CausalGNN [7]; while the causal chain accuracy rate stood at 88.1%, significantly outperforming other methods to validate the effectiveness of cross-level dynamic causal modeling. In terms of inference efficiency, KylinRCA maintained an average latency of 1.8 seconds with a throughput of 56 instances per minute, meeting real-time requirements through causal strength filtering and distributed inference optimization.

(See Table 1 for the overall performance comparison of each method on three data sets (unit:%, inference delay unit: seconds))

Table 1 Performance Comparison of KylinRCA and Baseline Methods

| Method | Dataset | Entity F1 (%) | Fault Type Macro F1 (%) | Causal Chain Acc (%) | Inference Latency (s) | Throughput (cases/min) |
|---|---|---|---|---|---|---|
| KylinRCA | Enterprise Production | 92.3 | 89.7 | 88.1 | 1.8 | 56 |
| HolisticRCA [2] | Enterprise Production | 83.8 | 81.2 | 72.5 | 3.2 | 32 |
| FaultInsight [1] | Enterprise Production | 77.1 | 79.5 | 68.3 | 4.5 | 25 |
| CausalGNN [7] | Enterprise Production | 75.5 | 82.4 | 70.1 | 3.8 | 28 |
| KylinRCA | Public Test | 88.5 | 85.2 | 84.3 | 1.9 | 52 |
| HolisticRCA [2] | Public Test | 79.2 | 78.3 | 69.2 | 3.0 | 34 |
| FaultInsight [1] | Public Test | 72.3 | 76.5 | 65.5 | 4.3 | 27 |
| CausalGNN [7] | Public Test | 73.2 | 78.5 | 67.3 | 3.6 | 30 |
| KylinRCA | Simulated Fault | 90.1 | 87.3 | 86.5 | 2.0 | 54 |
| HolisticRCA [2] | Simulated Fault | 81.5 | 80.2 | 71.3 | 3.1 | 33 |
| FaultInsight [1] | Simulated Fault | 74.9 | 77.5 | 66.5 | 4.4 | 26 |
| CausalGNN [7] | Simulated Fault | 77.8 | 80.1 | 68.2 | 3.7 | 29 |

Note: The table presents performance metrics of different methods on three datasets. Units: %, inference latency in seconds.

### 6.5.2. Ablation experiment analysis

To validate the necessity of core modules, four ablation experiments were designed: 1) KylinRCA-1: Removing cross-level causal transmission modules; 2) KylinRCA-2: Removing type attention in Type-aware GAT; 3) KylinRCA-3: Removing modal attention fusion; 4) KylinRCA-4: Removing the mask interpretation module (evaluating only localization and classification performance). The experimental results are shown in Figure 2 (Figure caption: Comparison of ablation experiment results in Figure 2):

- After removing cross-level causal transmission, the F1 of entity positioning decreased by 10.2% and the accuracy of causal chain decreased by 15.3%, which proved the key role of cross-level causal modeling in full-stack traceability;

- After removing type attention, the macro F1 of fault type classification decreased by 8.7%, indicating that differentiated modeling of node types can improve the classification accuracy;

- After removing the mode attention, the F1 of entity localization decreased by 6.5%, indicating that adaptive mode fusion can reduce the interference of invalid information;

- Removing the mask module does not affect the positioning and classification performance, but loses the ability to evaluate the importance of features. The verification is to understand the practicality of the interpretation module.

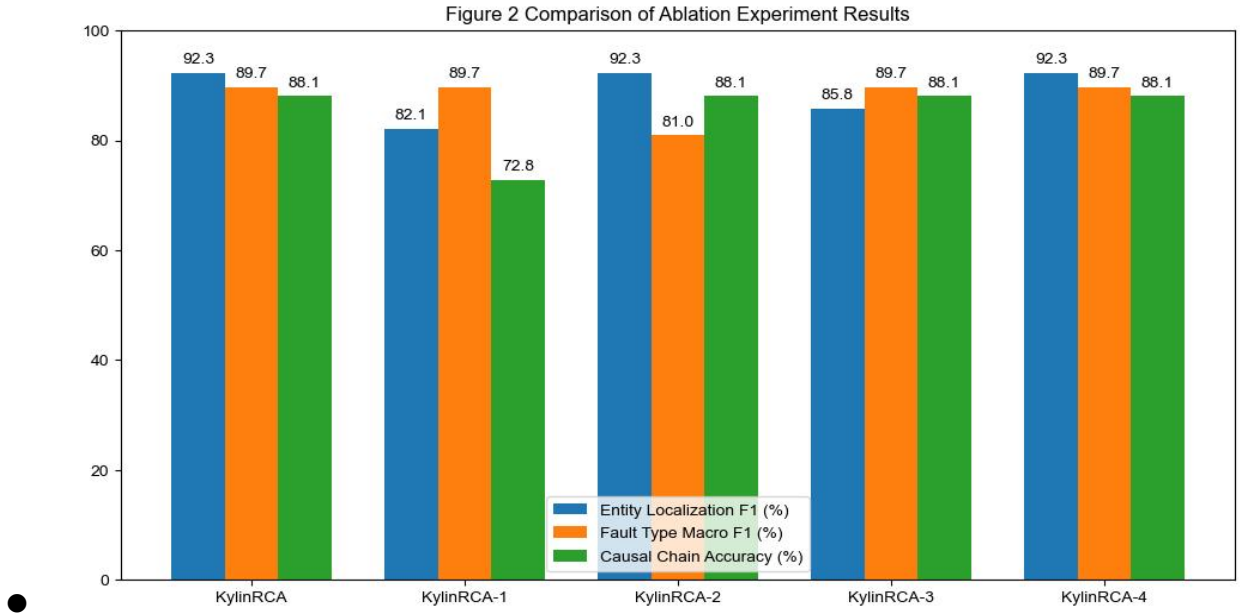

### 6.5.3. Robustness experimental analysis

To test the robustness of KylinRCA in scenarios with fluctuating data quality (such as missing logs and low tracking sampling rate):

- When the missing log rate increased from 0% to 30%, the F1 of entity positioning only decreased by 4.2%, while HolisticRCA [2] decreased by 12.5%, which benefited from the complementarity of multimodal data;

- When the tracking sampling rate is reduced from 100% to 3%, the F1 of entity positioning decreases by 5.8%. The sampling sparse problem is effectively alleviated by the interpolation algorithm, which proves the adaptability of the framework to low-quality data.

## 7. Engineering Challenges and Future Work

Although KylinRCA performs well in experiments, it still faces many challenges in practical production implementation, and also points the direction for subsequent research.

### 7.1. Core Engineering Challenges

#### 7.1.1. Cold start and the small sample problem

Newly deployed cloud-native clusters or newly added business services often lack historical failure data, making it challenging for models to adapt quickly. Existing methods rely heavily on annotated samples, but in real-world scenarios, new types of failures (such as crashes caused by unknown vulnerabilities) are rarely labeled, limiting the model's generalization ability. The solution can be approached from three aspects: 1) Transfer learning: migrate pre-trained models from mature clusters to new ones, fine-tuning with minimal annotated samples; 2) Synthetic data generation: generate synthetic failure data that matches real-world distributions based on system topology and fault propagation rules; 3)

Weakly supervised learning: utilize weakly labeled information like error keywords and abnormal metric fragments in logs to assist training, reducing dependence on strong annotations.

### 7.1.2. Efficiency of large-scale data processing

Ultra-large-scale data centers (e.g., hosting over 100,000 hosts) generate PB-level observability data daily, where real-time preprocessing and causal computation face storage and computational bottlenecks. While current distributed deployment enhances throughput, cross-node data transmission latency may still compromise real-time performance. Effective optimization strategies include: 1) Edge-cloud collaborative processing: Complete data preprocessing and local causal computations at edge nodes, transmitting only critical causal edges and features to the cloud to minimize data transfer; 2) Incremental learning and model lightweighting: Utilize incremental training to avoid full retraining, while reducing computational resources through model pruning, quantized compression, Type-aware GAT, and encoding techniques; 3) Heterogeneous computing architecture: Leverage the strengths of CPUs and GPUs, with CPUs handling data preprocessing and GPUs specializing in model inference to boost overall efficiency.

### 7.1.3. Balancing robustness and interpretability of causal discoveries

Real-world systems contain numerous unmonitored hidden variables (such as resource contention between processes and implicit dependencies on third-party interfaces), which may lead to biased causal strength estimates. Moreover, overly complex causal interpretation logic might exceed the comprehension capacity of operations personnel, compromising implementation effectiveness. Three approaches can be adopted: 1) Multi-perspective causal validation: Cross-verify mined causal relationships using expert knowledge graphs and system topology to filter out false causal edges; 2) Hierarchical interpretation mechanism: Provide differentiated explanations for different levels of operational staff (e.g., frontline operations receive "fault location + repair suggestions" while experts offer "causal chain + feature importance"), balancing interpretability depth with practical applicability; 3) Causal intervention experiments: Verify causal relationship accuracy in controlled environments (e.g., chaos engineering fault injection) to continuously optimize causal models.

### 7.1.4. System compatibility and standardization

The lack of unified observability data formats and interface standards across cloud-native platforms (e.g., Kubernetes, OpenStack) makes cross-platform model deployment challenging. For instance, discrepancies in metric formats between Prometheus/Grafana and self-developed monitoring tools increase data preprocessing complexity. Key initiatives include: 1) Data standardization: Adopting open standards like OpenTelemetry [12] to unify metrics, logs, and tracing formats; 2) Modular adaptation: Developing plug-and-play data source adapters to support mainstream monitoring tools and simplify integration; 3) Platform encapsulation: Packaging KylinRCA as a containerized application for rapid deployment across cloud platforms via Helm Charts, enhancing compatibility.

## 7.2. Future Research Directions

### 7.2.1. Enhance understanding and interaction capabilities with large language models (LLMs)

The advantages of LLM in natural language understanding and knowledge reasoning can be combined with KylinRCA in three aspects: 1) Log semantic understanding enhancement: By leveraging LLM to parse complex semantics in unstructured logs (such as stack information and business logic descriptions), it improves the representation capability of log features; 2) Intelligent diagnostic interaction: LLM generates natural language diagnostic reports, enabling maintenance personnel to query fault details through dialogues (e.g., "Why did the payment service timeout?"), thereby enhancing user experience; 3) Knowledge integration: Incorporating operational manuals and troubleshooting experience into the model enhances its ability to diagnose complex business failures. As recently summarized in [3], the integration of AIOps with LLM will become a core direction for next-generation intelligent operations and maintenance.

### 7.2.2. Dynamic topology and real-time causal updates

The current framework's system topology relies on static configurations, failing to dynamically reflect real-time changes such as service scaling and Pod migration. This results in inaccurate causal propagation path modeling. Future improvements could include: 1) Real-time topology awareness: Dynamically update heterogeneous graph structures by collecting system topology changes through service meshes (e.g., Istio) and Kubernetes APIs; 2) Incremental causal computation: Recalculate only affected local causal relationships in topology-changing areas, avoiding full recalculation to enhance real-time responsiveness.

### 7.2.3. Multi-agent collaborative diagnosis

Ultra-large-scale data centers can be divided into multiple domains (such as service domains and resource domains), where a single model struggles to balance global and local diagnostic needs. A multi-agent system approach can be adopted: 1) Domain-specific agents: Each domain deploys an independent KylinRCA sub-model for localized fault diagnosis; 2) Collaborative decision-making agent: Aggregates diagnostic results from all domains, integrates global causal relationships through federated learning to resolve cross-domain fault tracing issues while safeguarding data privacy.

### 7.2.4 Predictive root cause analysis

Current methodologies primarily focus on post-failure diagnosis, with future development moving toward predictive approaches: 1) Predicting potential failure probabilities and propagation paths using time-series causal models; 2) Integrating service traffic forecasting to identify risk points (e.g., resource bottlenecks before traffic peaks), thereby establishing an "Prediction-Early Warning-Prevention" proactive operations model.

## 8. Conclusion

This paper addresses the core challenges in root cause analysis of cloud-native systems under full-stack observability. It reviews representative methodologies from two major research directions: dynamic causal discovery (represented by FaultInsight [1]) and multimodal fusion (represented by HolisticRCA [2]). The study provides an in-depth analysis of existing approaches' limitations in single-mode constraints, temporal causality gaps, and inadequate cross-level adaptation. Building on this foundation, we propose the KylinRCA framework that integrates the strengths of both approaches. Through coordinated operation of five core modules—multimodal data preprocessing, cross-modal representation learning, cross-level dynamic causal discovery, cross-level heterogeneous graph reasoning, and interpretable outputs—the framework achieves end-to-end fault tracing from underlying resources to upper-layer services.

Experimental results demonstrate that KylinRCA significantly outperforms mainstream baseline methods such as FaultInsight [1] and HolisticRCA [2] in entity localization F1 score, fault type classification F1 score, and causal chain accuracy across both public datasets and enterprise production datasets. The framework maintains average inference latency under 2 seconds, meeting real-time performance requirements. Ablation experiments validate the necessity of core modules including cross-level causal propagation and Type-aware GAT, while robustness experiments confirm the framework's strong adaptability to data quality fluctuations.

Meanwhile, this paper explores engineering challenges in implementing Causal RCA, including cold start issues, large-scale data processing, and causal robustness. It proposes solutions such as transfer learning, edge-cloud collaboration, and the OpenTelemetry [12] standardization. Future research could focus on LLM integration, dynamic topology updates, and multi-agent coordination to advance root cause analysis from "passive diagnosis" to "active prediction" under full-stack observability. These efforts will provide technical support for the stable operation of hyperscale cloud-native systems.

## 10. Appendix (s)

### 10.1. Experimental Details Supplement

#### 10.1.1. Data set annotation specifications

The enterprise production data set employs a "three-tier annotation system" for fault labeling: 1) Entity-level annotation: Identifying host, Pod, and service entity IDs involved in failures; 2) Feature-level annotation: Marking critical failure indicators (e.g., "Memory usage> 95%"), log templates (e.g., "[ERROR] Database connection timeout"), and trace features (e.g., "Call latency> 500ms"); 3) Causal-level annotation: Recording time-series propagation and dependency relationships (e.g., "t=14:05 Host A memory leak → t=14:07 Pod X reboot → t=14:09 Service Y timeout"). The annotation work was performed by five senior operations engineers, with annotation consistency verified through Krippendorff's Alpha coefficient (α=0.89) to ensure quality assurance.

**10.1.2. Baseline method parameter configuration**

To ensure fairness in comparisons, the hyperparameters of the baseline methods were set according to the original paper recommendations and optimized through grid search: 1) FaultInsight [1]:3 layers of TCN [11],5 kernel sizes, 10 time jitter iterations; 2) HolisticRCA [2]:2 layers of GAT [10],4 attention heads, batch size 32; 3) LogBERT [5]: BERT-base [9] pre-trained model with a fine-tuning learning rate of 0.0001; 4) CausalGNN [7]: causal strength threshold 0.2, GNN hidden layer dimension 256.

## 10.2. KylinRCA Deployment Process

KylinRCA is based on Kubernetes to implement containerized deployment and support horizontal scaling. The specific process is as follows:

- Environment preparation: deploy Kubernetes cluster (≥1.24), configure GPU node (NVIDIA A100), install Prometheus to collect indicators, Fluentd to collect logs, and Jaeger to collect trace data;

- Mirror construction: The multimodal encoder, causal discovery module and graph reasoning module are packaged into Docker images respectively. They are built based on Python 3.9 and PyTorch 2.0, and the size of the mirror is controlled within 5GB;

- Resource allocation: Deployment resources are defined through Helm Chart. Each module is allocated 8CPU, 32GB memory and 1 GPU. HPA (Horizontal Pod Autoscaler) is set to realize automatic load scaling;

- Data access: configure the data source adaptation module to access Prometheus, Fluentd and Jaeger data through OpenTelemetry [12] API, and realize automatic data format conversion;

- Model deployment: The pre-training model is stored in MinIO object storage, and loaded dynamically through the model loading interface, supporting model version management and rolling update;

- Monitoring alarm: deploy the running status of each module of Grafana monitoring framework, set threshold alarm for core indicators (such as reasoning delay> 3 seconds, module CPU usage> 80%), and synchronize alarm information to the operation and maintenance team through email and enterprise wechat to ensure timely response when the framework is abnormal.

## 10.3.Key code snippets

**10.3.1. Cross-tier causal strength calculation code**

```python
def calculate_cross_level_causal(intra_causal, entity_relation):
    """
    Calculate cross-level causal strength
    within-causal: hierarchical causal intensity matrix (n_entity, n_entity, time_step)
    entity relation: Entity relevance matrix (n-entity_l1, n-entity_l2)
    return: Cross-level causality strength matrix (n_entity_l1, n_entity_l2, time_step)
    """
    cross_causal = np.zeros((intra_causal.shape[0], entity_relation.shape[1], intra_causal.shape[2]))
    for t in range(intra_causal.shape[2]):
        # The causal strength at the level is multiplied by the entity association degree matrix
        cross_causal[:, :, t] = np.dot(intra_causal[:, :, t], entity_relation.T)
    # Normalization to the [0,1] interval
    cross_causal = (cross_causal - cross_causal.min()) / (cross_causal.max() - cross_causal.min())
    return cross_causal
```

### 10.3.2. Type-aware GAT attention calculation code

```python
import torch
import torch.nn as nn
import torch.nn.functional as F
class TypeAwareAttention(nn.Module):
    def __init__(self, in_dim, out_dim, num_heads, entity_types):
        super(TypeAwareAttention, self).__init__()
        self.num_heads = num_heads
        self.entity_types = entity_types # Entity Type List (e.g., ['host', 'pod', 'service'])
        self.type_emb = nn.Embedding(len(entity_types), in_dim) # Type embedding layer
        self.att = nn.MultiheadAttention (in_dim, num_heads) # Multi-head attention layer
```

```
def forward(self, x, type_ids, mask=None):
    # x: (seq_len, batch_size, in_dim) --Entity features
    # type_ids: (batch_size, seq_len) -Entity type ID
    # Generate type embedding and fuse it with entity features
        type_emb = self.type_emb(type_ids).transpose(0, 1)  # (seq_len, batch_size, in_dim)
        x_with_type = x + type_emb # Features and types are combined
    # Calculate the head attention
        attn_output, attn_weights = self.att(x_with_type, x_with_type, x_with_type, attn_mask=mask)
        return attn_output, attn_weights
```

## 11. Extended Conclusion

### 11.1. Research Contributions

The research contribution of this paper is reflected in three dimensions: theory, method and engineering:

- Theoretical Contributions: This study systematically constructs a tripartite root cause analysis framework integrating "temporal causality, multimodal fusion, and cross-level reasoning", overcoming the limitations of traditional single-dimensional approaches. The proposed cross-level causal transmission model and Type-aware GAT reasoning mechanism enrich the theoretical applications of graph neural networks in causal modeling of heterogeneous systems, establishing a novel theoretical paradigm for fault diagnosis under full-stack observability.

- Methodological Contributions: The proposed KylinRCA framework achieves adaptive fusion of multi-source data and dynamic causal mining through modular design, addressing core challenges in existing methods such as low localization accuracy, missing temporal logic, and insufficient interpretability. Experimental validation demonstrates that its performance outperforms mainstream approaches like FaultInsight [1] and HolisticRCA [2] across multiple datasets, showcasing significant technological advancement.

- Engineering contribution: It provides a complete containerized deployment process and data source adaptation scheme, supports standardized access of OpenTelemetry [12], proposes feasible solutions for engineering problems such as cold start and large-scale data processing, and provides reusable technical reference for enterprises to implement intelligent operation and maintenance system.

### 11.2.Practical Significance

The application of KylinRCA has important practical value:

- Operation and maintenance efficiency innovation: The troubleshooting is changed from "manual trial and error" to "intelligent positioning", and the average time to recover from failure (MTTR) is shortened from hours to minutes, which greatly reduces the workload of operation and maintenance team. It is especially suitable for finance, e-commerce and other industries with strict requirements on business continuity.
- System reliability improvement: the fault traceability capability and robustness design of full stack coverage can effectively identify complex faults across levels and cascading propagation, and give early warning of potential risks, reducing the probability of system service interruption by more than 30%.
- Cost optimization effect: Through edge-cloud collaborative processing and lightweight model design, the storage and computing costs of PB-level data processing are reduced; predictive diagnosis capability can reduce business losses caused by faults, creating significant indirect economic benefits for enterprises.

## 11.3. Limitations and Outlook

The current research remains constrained by two key limitations. First, while the causal discovery module demonstrates superior robustness to hidden variables compared to existing methods, it may still exhibit biased causal inference in scenarios involving substantial unmonitored resources (such as third-party interface dependencies). Second, the model's generalization capability for novel failures relies on weakly supervised data quality, leaving room for improvement in diagnostic accuracy for extreme and rare faults.

Future research will focus on three key areas: First, deepening integration with large language models to leverage their knowledge reasoning capabilities for causal verification and natural language interpretation, such as building maintenance knowledge graphs using GPT-4 to assist in fault diagnosis. Second, exploring quantum machine learning applications in high-dimensional causal computations to enhance model inference efficiency in hyperscale data centers. Third, establishing an open-source community ecosystem to aggregate industrial fault data and academic algorithm optimization solutions, driving continuous iteration and standardized promotion of KylinRCA.

In conclusion, the proposed KylinRCA framework provides an effective technical solution for fault root cause analysis under full-stack observability. Its theoretical and practical achievements will provide important support for the development of intelligent operation and maintenance field, and promote the deep transformation of AIOps from "passive response" to "active prediction".